\documentclass[11pt,a4paper]{article}

\usepackage[left=3cm,right=3cm,top=3cm,bottom=4cm]{geometry}
\usepackage[usenames,dvipsnames,table]{xcolor}
\usepackage[super,nospace]{cite}
\usepackage{algpseudocode}
\usepackage{algorithm}
\usepackage{amssymb}
\usepackage{amsmath}
\usepackage{booktabs}
\usepackage{multirow}
\usepackage{graphicx}
\usepackage{hyperref}
\usepackage{fullpage}
\usepackage{caption}
\usepackage{lettrine} 
\usepackage{titlesec}
\usepackage{xspace}
\usepackage{lipsum}
\usepackage{subfig}
\usepackage{times}
\usepackage{cases}
\usepackage{array}
\usepackage{tikz}

\DeclareGraphicsExtensions{.eps,.pdf,.jpg,.png}

\newenvironment{intro}{\noindent\ignorespaces}{\par\noindent\ignorespacesafterend}

\makeatletter
\renewcommand\@biblabel[1]{#1.}
\makeatother

\definecolor{black}{RGB}{0, 0, 0}

\newcommand{\col}[1]{``#1''}

\newcommand{\mean}[1]{\mu_{#1}}
\newcommand{\avg}[1]{\left<#1\right>}
\newcommand{\std}[1]{\sigma_{#1}}
\newcommand{\esr}[1]{\hat{#1}}
\newcommand{\und}[1]{#1'}

\newcommand{\hyp}[0]{H}
\newcommand{\nhyp}[0]{\hyp_0}

\newcommand{\pval}[0]{P\mbox{-value}}

\newcommand{\pcc}[0]{r}
\newcommand{\scc}[0]{\rho}

\newcommand{\nets}[0]{N}
\newcommand{\stats}[0]{K}
\newcommand{\rank}[0]{R}
\newcommand{\cv}[0]{q}

\newcommand{\G}[0]{G}
\newcommand{\V}[0]{V}
\newcommand{\E}[0]{L}
\newcommand{\WCC}[0]{WCC\xspace}

\newcommand{\n}[0]{n}

\newcommand{\m}[0]{m}

\newcommand{\x}[0]{x}
\newcommand{\kd}[0]{k}
\newcommand{\ki}[0]{\kd_{in}}
\newcommand{\ko}[0]{\kd_{out}}
\newcommand{\kmin}[0]{\kd_{min}}
\newcommand{\h}[0]{\Delta}

\newcommand{\plc}[0]{\gamma}
\newcommand{\plci}[0]{\plc_{in}}
\newcommand{\plco}[0]{\plc_{out}}
\newcommand{\trian}[0]{t}
\newcommand{\omg}[0]{\omega}
\newcommand{\mix}[0]{r}
\newcommand{\mixii}[0]{\mix_{(in,in)}}
\newcommand{\mixio}[0]{\mix_{(in,out)}}
\newcommand{\mixoi}[0]{\mix_{(out,in)}}
\newcommand{\mixoo}[0]{\mix_{(out,out)}}
\newcommand{\clus}[0]{c}
\newcommand{\bclus}[0]{b}
\newcommand{\dclus}[0]{d}
\newcommand{\mixc}[0]{\mix_{\clus}}
\newcommand{\mixb}[0]{\mix_{\bclus}}
\newcommand{\mixd}[0]{\mix_{\dclus}}
\newcommand{\diam}[0]{\delta}
\newcommand{\ediam}[0]{\diam_{90}}

\newcommand{\C}[1]{C(#1)}
\newcommand{\B}[1]{B(#1)}
\newcommand{\D}[1]{D(#1)}
\newcommand{\N}[1]{N(#1)}
\newcommand{\HP}[1]{H(#1)}

\algrenewcommand{\algorithmiccomment}[1]{\hfill\{#1\}}

\newcommand{\resref}[0]{Results\xspace}
\newcommand{\disref}[0]{Discussion\xspace}
\newcommand{\metref}[0]{Methods\xspace}
\newcommand{\sfigref}[1]{Figure~\ref{fig:#1}\xspace}
\newcommand{\figref}[1]{Fig.~\ref{fig:#1}\xspace}
\newcommand{\tblref}[1]{Table~\ref{tbl:#1}\xspace} 
\renewcommand{\eqref}[1]{equation~(\ref{eq:#1})\xspace}

\newcommand{\wos}[0]{{\itshape Web of Science}\xspace}
\newcommand{\scp}[0]{{\itshape Scopus}\xspace}
\newcommand{\cs}[0]{{\itshape CiteSeer}\xspace}
\newcommand{\csx}[0]{{\itshape CiteSeer$^{\mathrm{x}}$}\xspace}
\newcommand{\gs}[0]{{\itshape Google Scholar}\xspace}
\newcommand{\hc}[0]{{\itshape Algorithmic Historiography}\xspace}
\newcommand{\cora}[0]{{\itshape Computer Science Research Paper Search Engine}\xspace}
\newcommand{\arxiv}[0]{{\itshape arXiv.org}\xspace}
\newcommand{\dblp}[0]{{\itshape DBLP Computer Science Bibliography}\xspace}
\newcommand{\gnutella}[0]{Gnutella peer-to-peer file sharing\xspace} 
\newcommand{\twitter}[0]{Twitter social circles\xspace} 

\newcommand{\WoS}[0]{WoS\xspace} 
\newcommand{\CS}[0]{CiteSeer\xspace} 
\newcommand{\HC}[0]{HistCite\xspace} 
\newcommand{\Cora}[0]{Cora\xspace} 
\newcommand{\arXiv}[0]{arXiv\xspace} 
\newcommand{\DBLP}[0]{DBLP\xspace} 
\newcommand{\Gnutella}[0]{Gnutella\xspace} 
\newcommand{\Twitter}[0]{Twitter\xspace}

\begin{document}


\title{Network-based statistical comparison of citation topology of bibliographic databases}
\author{Lovro \v{S}ubelj$^{1,*}$, Dalibor Fiala$^{2}$ \& Marko Bajec$^{1}$}
\date{} 

\maketitle

\centerline{{\footnotesize $^{1}$ University of Ljubljana, Faculty of Computer and Information Science, 
    Ve\v{c}na pot 113, SI-1000 Ljubljana, Slovenia}}
\centerline{{\footnotesize $^{2}$ University of West Bohemia, Faculty of Applied Sciences, Univerzitn\'{i} 8, CZ-30614 Plze\u{n}, Czech Republic}}
\centerline{{\footnotesize $^*$ Corresponding author: lovro.subelj@fri.uni-lj.si}}


\abstract{Modern bibliographic databases provide the basis for scientific research and its evaluation. While their content and structure differ substantially, there exist only informal notions on their reliability. Here we compare the topological consistency of citation networks extracted from six popular bibliographic databases including \wos, \cs and \arxiv. The networks are assessed through a rich set of local and global graph statistics. We first reveal statistically significant inconsistencies between some of the databases with respect to individual statistics. For example, the introduced field bow-tie decomposition of \dblp substantially differs from the rest due to the coverage of the database, while the citation information within \arxiv is the most exhaustive. Finally, we compare the databases over multiple graph statistics using the critical difference diagram. The citation topology of \dblp is the least consistent with the rest, while, not surprisingly, \wos is significantly more reliable from the perspective of consistency. This work can serve either as a reference for scholars in bibliometrics and scientometrics or a scientific evaluation guideline for governments and research agencies.}


\section*{\label{sec:intro}}

\begin{intro}
  \lettrine[lines=3,slope=-3pt,findent=-2pt]{B}{}ibliographic databases range from expensive hand-curated professional solutions like \wos and \scp to preprint repositories~\cite{Gin11}, public servers~\cite{Ley02} and automated services that collect freely accessible manu\-scripts from the Web~\cite{BLG98,MNRS00}. These provide the basis for scientific research, where new knowledge is derived from the existing, while also the main source of its evaluation. Undoubtedly, the number of citations a paper receives is still considered to be the main indicator of its importance or relevance~\cite{WSB13,New14}. However, the probability distribution of scientific citations has been shown to follow a wide range of different forms including power-law~\cite{Pri65}, shifted power-law~\cite{EF11}, stretched exponential~\cite{LS98}, log-normal~\cite{RFC08}, Tsallis~\cite{WLG09}, and modified Bessel~\cite{Van01}, to name just a few. Although some methods used in these studies might be questionable, more importantly, they are based on different bibliographic data. In fact, the content and structure of modern bibliographic databases differ substantially, while there exist only informal notions on their reliability. 
\end{intro}

One way to assess the databases is simply by the amount of literature they cover. \wos spans over $100$ years and includes several dozens of millions of publication records~\cite{Gar55,Fia11}, an extent similar to that of \scp, which, however, came into existence only some ten years ago. On the other hand, the preprint repository \arxiv~\cite{Gin11} and the digital library \dblp~\cite{Ley02} both date back to $1990$s and include only millions of publications or publication records. The coverage of different bibliographic databases has else been investigated by various scholars~\cite{FPMP08,VG09,Fia11,DR12b}, while others have analyzed also their temporal evolution~\cite{Red05,Gin11}, available features~\cite{Jac05,FPMP08}, data acquisition and maintenance methodology~\cite{PCHCG05,Fia11}, and the use within a typical scientific~workflow~\cite{HPK08}.

Yet, despite some notable differences, the reliability of bibliographic databases is primarily seen as the accuracy of its citation information. While citations are input by hand in the case of professional databases, services like \cs and \gs use information retrieval and machine learning techniques to automatically parse citations from publication manuscripts~\cite{BLG98,MNRS00}. Expectedly, this greatly impacts bibliometric analyses~\cite{PCHCG05} and standard metrics of scientific evaluation like citation counts and $h$-index~\cite{MY07,DR12b}. Although networks of citations between scientific papers have been studied since the $1950$s~\cite{Gar55,Pri65}, and are also commonly used in the modern network analysis literature~\cite{LKF07,LSB11b}, there exists no statistical comparison of citation topology of different bibliographic databases.

In this study, we compare the topological consistency of citation networks extracted from six popular bibliographic databases (see~\metref). The networks are assessed through local and global graph statistics by a methodology borrowed from the machine learning literature~\cite{Dem06}. We first reveal statistically significant inconsistencies between some of the databases with respect to individual graph statistics. For example, the introduced field bow-tie decomposition of \dblp substantially differs from the rest due to the coverage of the database or the sampling procedure, while the citation information within \arxiv is proven to be the most exhaustive. Finally, we compare the consistency of databases over multiple graph statistics. The citation topology of \dblp is the least consistent with the rest, while, not surprisingly, \wos is significantly more reliable from this perspective. Note that the reliability is here seen as a deviation from the majority (see~\disref). Differences between other databases are not statistically significant. This work can serve either as a reference for scholars in bibliometrics and scientometrics or a scientific evaluation guideline for governments and~research~agencies.


\titleformat*{\paragraph}{\small\bf}

\section*{\label{sec:results}Results}

Citation networks representing bibliographic databases are compared through $21$ graph statistics described in~\metref. In the following, we discuss the values of statistics in the context of complex network theory. Next, we reveal some statistically significant differences in individual statistics using Student $t$-test~\cite{CW82}. We then select ten statistics whose independence is confirmed by Fisher $z$-test~\cite{Fis15} and show that the databases display significant inconsistencies in the selected statistics using Friedman rank test~\cite{Fri37,Fri40}. Last, the databases with no significant inconsistencies are revealed by Nemenyi post-hoc test~\cite{Nem63} and the critical difference diagram~\cite{Dem06}. Finally, we also compare the bibliographic databases with the selected online databases to verify the predictive power of the employed statistical methodology. See~\metref for further details on statistical~comparison.

\paragraph{Graph statistics of citation networks.} 

\tblref{statistics} shows descriptive statistics of citation networks. The networks range from thousands of nodes to millions of links, while the largest weakly connected components contain almost all the nodes. This is consistent with the occurrence of a giant connected component in random graphs~\cite{ER59}. Directed networks are often assessed also according to their bow-tie structure~\cite{BKMRRSTW00}. However, due to the acyclic nature of citation networks where papers can only cite papers from the past, the decomposition proves meaningless. We introduce the field bow-tie decomposition into the in-field component, which consists of papers citing no other paper, the out-field component, which consists of papers not cited by any other paper, and the field core. The out-field component thus includes the research front~\cite{Pri65}, and the in-field and core components include the knowledge or intellectual base~\cite{Per94}. \tblref{statistics} shows the percentage of nodes in each of the field components, while a visual representation is given in~\figref{profile}. Notice that, in most cases, the majority of papers is included in the core and out-field components of the citation networks. Nevertheless, the main mass of the papers shifts towards the in-component in \HC and \DBLP databases (\sfigref{profile}, panels {\bf D} and {\bf E}). Since the former consists of papers from merely major journals and conferences, and the latter is based on the bibliography of a single author, many of the papers in the databases cite no other. Hence, reducing a bibliographic database to only a subset of publications or authors gives notably different citation structure and also influences many common graph statistics.

\tblref{degree} shows degree statistics of citation networks. Observe that the mean degree $\avg{\kd}$ is around $8.8$ in all cases except \arXiv database, which, somewhat surprisingly, coincides with the common density of real-world networks~\cite{LJTBH11}. Note, however, that since $\avg{\kd}/2=\avg{\ki}=\avg{\ko}$ for any network, the papers cite and are cited by only four other papers on average. This number becomes meaningful when one considers that far more citations come from outside the field~\cite{Red04,Red05}, whereas all databases are subsets of their respective fields in some sense. Considerably higher $\avg{\kd}$ in \arXiv database is most likely due to several reasons. In contrast to other databases, \arxiv stores journal and conference papers, technical reports, draft manuscripts that never came to print etc. Next, the citation network studied has been released within the KDD Cup $2003$ (\href{http://www.cs.cornell.edu/projects/kddcup}{{\scriptsize\tt http://www.cs.cornell.edu/projects/kddcup}}) and has thus presumably been cleansed appropriately. Also, the subset of \arxiv considered consists of physics publications, while other databases consist of computer science publications. Regardless of the true reason, the citation information within \arXiv database is notably more exhaustive, which clearly reflects in its graph structure (see~field bow-tie in \figref{profile}, panel {\bf F}).

\sfigref{profile} plots degree distributions of citation networks, while the corresponding scale-free~\cite{BA99} exponents $\plc$, $\plci$ and $\plco$ are given in~\tblref{degree}. We stress that not all distributions, especially out-degree distributions, are a valid fit to a power-law form~\cite{CSN09}. Nevertheless, the degree distributions further confirm the inconsistencies observed above. A larger number of non-citing papers results in a less steep out-degree distribution, whereas $\plco\approx 2.6$ for \HC and \DBLP databases, while $\plco\approx 3.8$ otherwise. On the contrary, the in-degree distribution of \HC database is much steeper with $\plci=3.5$, while $\plci\approx 2.5$ for the rest. In fact, $\plci>\plco$ for \HC database, whereas $\plci<\plco$ for all others. Finally, the lack of low-citing papers in \arXiv database prolongs the degree distributions towards the right-hand side of the scale (see \figref{profile}, panel {\bf F}). 

Degree mixing~\cite{New02} in~\tblref{degree} reveals no particularly strong correlations. Still, the in-degree and out-degree mixing coefficients $\mixii$ and $\mixoo$ show positive correlation, while the undirected degree mixing $\mix$ is negative. For comparison, $\mix\gg 0$ in social networks, and $\mix\ll 0$ for Internet and the Web~\cite{New02,New03a}. Again, \HC and \DBLP databases deviate from common behaviour due to the reasons given above. For example, the directed degree mixing coefficient $\mixoi$ is substantially lower for \HC database, while all directed coefficients are relatively low for \DBLP database. \sfigref{profile} plots also neighbour connectivity profiles of citation networks. Notice dichotomous degree mixing~\cite{HL11} that is positive for smaller out-degrees and negative for larger in-degrees, represented by increasing or decreasing trend, respectively (see, e.g., \figref{profile}, panels {\bf A} and {\bf B}). Similar observations were recently made also in software~\cite{SZBB14} and undirected biological~\cite{HL11} networks. Consistent with the above, these trends are not present in \HC and \DBLP databases (see \figref{profile}, panels {\bf D} and~{\bf E}).

\tblref{clustering} shows clustering~\cite{WS98} statistics of citation networks. The mean clustering coefficients $\avg{\clus}$, $\avg{\bclus}$ and $\avg{\dclus}$ greatly vary across the databases, whereas $\avg{\clus}\approx 0.15$ for \WoS, \CS and \DBLP databases, and $\avg{\clus}\approx 0.3$ in the case of \Cora, \HC and \arXiv databases. This may be an artefact of the coverage or the sampling procedure used for citation extraction, while clustering can also reflect the amount of citations copied from other papers~\cite{SR03,SZB14} known as indirect citation~\cite{PPD10}. Unbiased clustering mixing coefficients $\mixb$ and $\mixd$ in~\tblref{clustering} reveal strong positive correlations, similar to other real-world networks~\cite{SZBB14}. However, as before, $\mixd=0.26$ for \DBLP database, while $\mixd\approx 0.4$ for all others. \sfigref{profile} plots clustering profiles of citation networks. Due to degree mixing biases~\cite{SV05}, $\C{\kd}\sim \kd^{-\alpha}$ for $\alpha\approx 1$~\cite{RB03}, while this behaviour is absent from corrected profiles $\B{\kd}$ and $\D{\kd}$.

\tblref{clustering} shows also diameter statistics of citation networks. Undirected effective diameter $\und{\ediam}$ is somewhat consistent across the databases, in contrast to the directed variant $\ediam$, where $\ediam\approx 8.5$ for \WoS, \HC and \DBLP databases, while $\ediam>20$ for other databases. Low value of $\ediam$ for \HC and \DBLP databases is due to the limited coverage discussed above, whereas the respective networks are also much smaller (see~\tblref{statistics}). On the other hand, low $\ediam$ for \WoS database is due to a rather non-intuitive phenomena that real-world networks shrink as they grow~\cite{LKF07}. \WoS database includes $50$ years of literature, while the time span of, e.g., \arXiv database is merely $10$ years. The databases are thus not directly comparable in $\ediam$ and neither is indeed inconsistent with the rest. Described can be more clearly observed in hop plots shown in \figref{profile} (see, e.g., panels {\bf A} and {\bf B}).

\paragraph{Comparison of databases by individual statistics.} 

The above discussion was in many cases~just qualitative. In the following, we reveal also statistically significant differences between some of the data\-bases with respect to individual graph statistics. Since their values of a {\it true} citation network are, obviously, not known, we compute externally studentized residuals that measure the consistency of each database with the rest (\sfigref{comparison}, panels {\bf A}-{\bf F}). Statistically significant inconsistencies in individual statistics are revealed by independent two-tailed Student $t$-tests (see~\metref).

\WoS, \CS and \Cora databases show no significant differences at $\pval=0.05$. On the contrary, the scale-free in-degree exponent $\plci$ in \HC database is significantly higher than in other databases, while the directed degree mixing coefficient $\mixoi$ is significantly lower ($\pval=0.019$ and $\pval=0.033$, respectively; see~\tblref{degree} and \figref{comparison}, panel {\bf D}). This is a direct consequence of the limited coverage already noted above. For example, since the database is derived from a bibliography of a single author, highly cited papers are likely missing, which results in a much steeper citation distribution $P(\ki)$ and thus higher $\plci$. Next, the unbiased clustering mixing coefficient $\mixd$ is significantly lower in \DBLP database ($\pval=0.017$; see~\tblref{clustering} and \figref{comparison}, panel~{\bf E}). Apparently, reducing the bibliographic database to only selected publications gives a rather heterogeneous citation structure, which does not share high clustering assortativity~\cite{SZBB14}, $\mixd\gg 0$, of other citation networks. Note that the differences in the field bow-tie decomposition of \DBLP database become statistically significant at $\pval=0.052$ (see below). Finally, as thoroughly discussed above, the citation information within \arXiv database is significantly more exhaustive with much higher mean degree $\avg{\kd}$ ($\pval=0.009$; see~\tblref{statistics} and \figref{comparison}, panel {\bf F}). Notice that statistically significant inconsistencies between the databases are, expectedly, merely a subset of the differences exposed through the expert analysis above. Still, in summary, the results reveal that bibliographic databases with substantially different coverage have significantly different citation topology.

At $\pval=0.1$, several other inconsistencies become statistically significant. For \CS database, the largest weakly connected component is significantly smaller than in other databases ($\pval=0.059$; see~\tblref{statistics} and \figref{comparison}, panel {\bf B}); for \HC database, the clustering mixing coefficient $\mixc$ is lower ($\pval=0.066$; see~\tblref{clustering} and \figref{comparison}, panel {\bf D}); for \DBLP database, the in-field component is larger ($\pval=0.052$; see~\tblref{statistics} and \figref{comparison}, panel {\bf E}), while the field core and the directed degree mixing coefficient $\mixii$ are smaller ($\pval=0.090$ and $\pval=0.095$, respectively; see~\tblref{statistics} and \tblref{degree}, and \figref{comparison}, panel {\bf E}); and for \arXiv database, the undirected degree mixing coefficient $\mix$ and the corrected clustering  coefficient $\avg{\bclus}$ are higher ($\pval=0.081$; see~\tblref{degree} and \tblref{clustering}, and \figref{comparison}, panel {\bf F}). Note that, due to space limitations, not all inconsistencies at $\pval=0.1$ are discussed in the analysis above.

\paragraph{Selection of independent graph statistics.} 

Since the adopted graph statistics of citation networks are by no means independent~\cite{WS98,SV05}, one cannot simply compare the bibliographic databases over all. For this purpose, we select ten statistics listed in~\figref{comparison}, panel {\bf G}, and verify their statistical independence (see~\metref). We compute Fisher transformations of the pairwise Spearman correlations between the statistics, while significant correlations are revealed by independent two-tailed $z$-tests (\sfigref{comparison}, panel {\bf H}). Notice that no correlation is statistically significant at $\pval=0.01$.

The selection of independent graph statistics proceeds as follows. We first discard statistics that are sums or aggregates of the others by definition. Namely, the sizes of the largest weakly connected and out-field components (see~\tblref{statistics}), the scale-free degree exponent $\plc$, the undirected degree mixing $\mix$ and also both mixed directed mixing coefficients $\mixio$ and $\mixoi$ (see~\tblref{degree}). We next discard statistics whose correlations have been proven in the literature~\cite{SV05} or are dependent on some intrinsic characteristic of the database like the time span of the publications (see above). Namely, the standard clustering $\avg{\clus}$ and the corresponding mixing coefficient $\mixc$, and the directed effective diameter $\ediam$ (see~\tblref{clustering}). Finally, out of the both unbiased clustering coefficients $\avg{\bclus}$ and $\avg{\dclus}$, we decide for the latter, and its corresponding mixing coefficient $\mixd$ (see~\tblref{clustering}). We are thus left with ten statistics (\sfigref{comparison}, panel {\bf G}). Namely, the sizes of the in-field and core components (see~\tblref{degree}), the mean degree $\avg{\kd}$, the directed scale-free exponents $\plci$ and $\plco$, and the directed degree mixing coefficients $\mixii$ and $\mixoo$ (see~\tblref{degree}), the unbiased clustering $\avg{\dclus}$ and its corresponding mixing coefficient $\mixd$, and the undirected effective diameter $\und{\ediam}$ (see~\tblref{clustering}).

For some further notes on statistics independence see~\disref.

\paragraph{Comparison of databases over multiple statistics.} 

In the following, we compare the bibliographic databases over independent graph statistics selected above. We rank the databases according to the studentized statistics residuals and compute their mean ranks over all statistics (see~\metref). The final ranks are $2.2$ for \WoS database, $3.1$ for both \CS and \Cora databases, $3.6$ for \arXiv database, $4.0$ for \HC database and $5.0$ for \DBLP database. Notice that the ranks indeed reflect the conclusions on database consistency given above. We reject the null hypothesis that the ranks of the databases are statistically equivalent by one-tailed Friedman test at $\pval=0.05$ and thus compare the ranks by two-tailed Nemenyi post-hoc test (\sfigref{comparison}, panel {\bf I}). The databases whose ranks differ by more than a critical distance $2.38$ show statistically significant inconsistencies in the selected statistics at $\pval=0.05$. Hence, the citation topology of \WoS database is significantly more reliable than that of \DBLP database, which is the least consistent with the rest. On the other hand, the differences between other databases are not statistically significant, whereas concluding that these are consistent with {\it both} \WoS and \DBLP databases would be a statistical nonsense~\cite{Dem06}. At $\pval=0.1$, the critical distance drops to $2.17$, while all conclusions still remain the same. Interestingly, neglecting the requirement for the independence of graph statistics and comparing the bibliographic databases over all $21$ statistics, again gives exactly the same conclusions on their consistency. Although, the ranking changes, since \arXiv database is ranked in front of \Cora~database.

For some further notes on database consistency see~\disref.

\paragraph{Comparison of bibliographic and online databases.} 

To assess the power of the employed statistical methodology for quantifying the differences in network topology, we compare citation networks representing different bibliographic databases with two networks extracted from online databases. Namely, a technological network of \gnutella (\href{http://rfc-gnutella.sourceforge.net}{{\scriptsize\tt http://rfc-gnutella.sourceforge.net}}) from August 2002~\cite{LKF07}, where nodes are hosts and links are shares between them; and a social network representing \twitter (\href{http://twitter.com}{{\scriptsize\tt http://twitter.com}}) crawled from public repositories~\cite{ML12d}, where nodes are users and links are follows between them. Both these networks are provided within SNAP (\href{http://snap.stanford.edu}{{\scriptsize\tt http://snap.stanford.edu}}), while their basic descriptive statistics are given in \tblref{statistics}.

Note that online databases reveal knowingly different network topology than reliable bibliographic databases. For example, the majority of nodes in \Gnutella database is included in the in-field component (see \metref), similarly as in \DBLP database (see \tblref{statistics}). Next, the mean degree $\avg{\kd}$ is considerably higher in \Twitter database and lower in \Gnutella database (see~\tblref{degree}). Furthermore, the degree distributions of \Gnutella database are not a valid fit to a power-law form~\cite{CSN09} with higher scale-free degree exponents $\plc$-s than in other databases (see \tblref{degree}). On the contrary, the scale-free out-degree exponent $\plco$ of \Twitter database is lower, similarly as in \HC database. Online databases also reveal notably different clustering regimes than bibliographic databases (see~\tblref{clustering}). The standard and unbiased clustering coefficients $\avg{\clus}$ and $\avg{\dclus}$ are much higher in \Twitter database, while much lower in \Gnutella database. Finally, \Gnutella database shows relatively heterogeneous clustering structure with very low unbiased clustering mixing coefficients~$\mixb$~and~$\mixd$.

In the following, we reveal statistically significant inconsistencies between some of the databases with respect to individual graph statistics (see \metref). We consider the online databases and four most reliable bibliographic databases so that all critical values remain the same as before. Under this setting, the bibliographic databases show no inconsistencies at $\pval=0.05$ (\sfigref{validation}, panels {\bf A}-{\bf D}). On the other hand, five most significant inconsistencies of online databases almost precisely coincide with the differences exposed through the analysis above (\sfigref{validation}, panels {\bf E} and {\bf F}). For \Gnutella database, the in-field component is larger ($\pval=0.008$), the degree and in-degree scale-free exponents $\plc$ and $\plci$ are higher ($\pval=0.011$ and $\pval=0.008$, respectively), and the unbiased clustering mixing coefficients $\mixb$ and $\mixd$ are lower ($\pval=0.032$ and $\pval=0.011$, respectively); and for \Twitter database, the mean degree $\avg{\kd}$ is higher ($\pval=0.039$), the out-degree scale-free exponent $\plco$ and the directed degree mixing coefficient $\mixii$ are lower ($\pval=0.063$ and $\pval=0.066$, respectively), and the standard and unbiased clustering coefficients $\avg{\clus}$ and $\avg{\dclus}$ are higher ($\pval=0.056$ and $\pval=0.065$, respectively).

In the remaining, we also rank the databases over multiple graph statistics as before (see \metref). We select ten statistics listed in~\figref{validation}, panel {\bf G}, whose pairwise independence is confirmed at $\pval=0.001$ (\sfigref{validation}, panel {\bf H}). The overall ranks of the databases are not statistically equivalent at $\pval=0.05$ and are given in~\figref{validation}, panel {\bf I}. Expectedly, the online databases are the least consistent with the rest, whereas the ranks are $4.6$ and $4.9$ for \Gnutella and \Twitter databases, respectively, and $1.9$--$3.3$ for the bibliographic databases. Yet, merely \WoS bibliographic database significantly differs from the online databases at $\pval=0.05$ (see~\figref{validation}, panel {\bf I}).

In summary, the employed statistical testing proves to be rather effective in quantifying the inconsistencies between network databases with respect to individual graph statistics. On the contrary, the comparison over multiple statistics appears to be less powerful and cannot distinguish between the online databases and all bibliographic databases considered above. Nevertheless, the statistically significant inconsistencies between \WoS and \DBLP bibliographic databases highlighted in the study can thus indeed be regarded as rather substantial.


\section*{\label{sec:discussion}Discussion}

We conduct an extensive statistical analysis of the citation information within six popular bibliographic databases. We extract citation networks and compare their topological consistency through a large number of graph statistics. We expose statistically significant inconsistencies between some of the databases with respect to individual graph statistics and compare the databases over multiple statistics. \dblp is found to be the least consistent with the rest, while \wos is significantly more reliable from this perspective. The result is somewhat surprising, since \dblp is informally considered as one of the most accurate freely available sources of computer science literature. The analysis further reveals that the coverage of the database and the time span of the literature greatly affect the overall citation topology, although this can be avoided in the case of the latter. This work can serve either as a reference for the analyses of citation networks in bibliometrics and scientometrics literature or a guideline for scientific evaluation based on some particular bibliographic database or literature coverage policy.

We introduce the field bow-tie decomposition of a citation network (see~\metref), which proves to be one of the most discriminative approaches for comparing the citation topology of bibliographic databases (see~\resref). We also consider $18$ other local and global graph statistics. Nevertheless, we neglect some possible common patterns of nodes like motifs~\cite{MSIKCA01} and graphlets~\cite{PWJ04}, and the occurrence of larger characteristic groups of nodes like communities~\cite{GN02} and modules~\cite{SB12d}. Yet, these structures are not well understood for the specific case of citations networks and thus not easily interpretable.

In the following, we provide some further notes on the representativeness and reliability of the bibliographic databases, and the independence of the databases and adopted graph statistics.

As discussed in~\metref, citation networks extracted from bibliographic databases are not necessarily representative due to citation retrieval procedure, data preprocessing techniques, size or other. It should, however, be noted that this work has been done after realizing that citation networks available from the Web provide a rather inconsistent view on the structure of bibliographic information. We have therefore collected and compared all such networks, while including also a citation network extracted from \wos. In that sense, the adopted networks are representative of the data readily available for the analyses and thus also commonly used in the literature~\cite{LKF07,LSB11b}. Still, other citation networks could give different conclusions on the reliability of bibliographic databases. In particular because the reliability is measured through consistency of the databases. The concepts are of course not equivalent, yet the study reveals that, in most cases, only a single database deviates from a common behaviour for some particular graph statistic (see~\resref). Hence, the reliability can indeed be seen as a deviation from the majority to a rather good approximation.

Independence between bibliographic databases is obtained trivially, since these are either based on independent bibliographic sources or cover different literature (see~\metref). On the other hand, adopted graph statistics of citation networks are by no means independent~\cite{WS98,SV05}. As this is required by several statistical tests, we reduce the statistics to a subset whose pairwise independence could be proven. Nevertheless, we only show that the statistics are not clearly dependent and we do not ensure their mutual independence. Although the conclusions of the study are exactly the same regardless of whether it is based on all or merely independent statistics (see~\resref), further reducing the subset of statistics would discard relevant information and no statistically significant conclusions could be made. We also stress that all results have been verified by an independent expert analysis. An alternative solution would be to transform the statistics into uncorrelated representatives using matrix factorization techniques like principal component analysis~\cite{Pea01}. 
However, interpreting inconsistencies in, e.g., $0.9\plci-1.4\mixc+0.3\ediam$ would most likely be far from trivial.


\titleformat*{\paragraph}{\footnotesize\bf}
\titleformat*{\subsection}{\normalsize\bf}

{\footnotesize

\section*{\label{sec:methods}Methods}

\subsection*{\label{sec:methods:data}Bibliographic sources}

In this study, we conduct a network-based comparison of citation topology of six bibliographic databases. These have been extracted from publicly available and commercial bibliographic sources, services, software and a preprint repository with particular focus on computer science publications. For bibliographic sources based on a similar methodology~\cite{FPMP08,Fia11} (e.g., \wos and \scp, \cs and \gs), a single exemplar has been selected. We have extracted a citation network from each of the selected databases. Publications neither citing nor cited by any other are discarded and any self-citations that occur due to errors in the databases are removed prior to the analysis (see below and~\tblref{statistics} for details). Although the databases contain fair portions of the respective bibliographic sources, we stress that they are not all necessarily representative. Still, in most cases, these are the only examples of citation networks readily available online (due to our knowledge) and thus also often used in the network~analysis~literature~\cite{LKF07,LSB11b}. 

\paragraph{\WoS database.} 

\wos (\WoS) is informally considered as the most accurate bibliographic source in the world. It is hand-maintained by professional staff at Thomson Reuters (\href{http://thomsonreuters.com}{{\scriptsize\tt http://thomsonreuters.com}}), previously Institute for Scientific Information. It dates back to the $1950$s~\cite{Gar55,Pri65} and contains over $45$ million~records of publications from all fields of science~\cite{Fia11}. For this study, we consider all journal papers in \WoS category {\it Computer Science, Artificial Intelligence} as of October $2013$. The extracted database spans $50$ years, and contains $179$,$510$ papers from $877$ journals and $639$,$126$ citations between them.
Note that $39$,$148$ papers neither cite nor are cited by any other, while the database includes $16$ self-citations.

\paragraph{\CS database.} 

\cs or \csx (\CS) is constructed by automatically crawling the Web for freely accessible manuscripts of publications and then analyzing the latter for potential citations to other publications~\cite{BLG98} (\href{http://citeseer.ist.psu.edu}{{\scriptsize\tt http://citeseer.ist.psu.edu}}). It became publicly available in $1998$ and is maintained by Pennsylvania State University. It contains over $32$ million publication records from computer and information science~\cite{Fia11}. For this study, we consider a snapshot of the database provided within KONECT (\href{http://konect.uni-koblenz.de}{{\scriptsize\tt http://konect.uni-koblenz.de}}) that contains $723$,$131$ publications and $1$,$751$,$492$ citations between~them.
Note that $338$,$718$ publications neither cite nor are cited by any other, while the database includes $6$,$873$~self-citations.

\paragraph{\Cora database.} 

\cora (\Cora) is a service for automatic retrieval of publication manuscripts from the Web using machine learning techniques~\cite{MNRS00} (\href{http://people.cs.umass.edu/~mccallum}{{\scriptsize\tt http://people.cs.umass.edu/\~{}mccallum}}). It contains over $50$,$000$ publication records collected from the websites of computer science departments at major universities in August $1998$. For this study, we consider a subset of the database that contains $23$,$166$ publications and $91$,$500$ citations between them~\cite{SB13} (\href{http://lovro.lpt.fri.uni-lj.si}{{\scriptsize\tt http://lovro.lpt.fri.uni-lj.si}}).
Note that all papers either cite or are cited by some other, while
the database includes no~self-citations.

\paragraph{\HC database.} 

\hc (\HC) is a software package for analysis and visualization of bibliographic databases owned by Thomson Reuters (\href{http://www.histcite.com}{{\scriptsize\tt http://www.histcite.com}}). It was developed in the $2000$s for extracting publication records from \WoS database~\cite{Gar04}. For this study, we consider a complete bibliography of Nobel laureate Joshua Lederberg produced by \HC in February $2008$. The database contains $8$,$843$ publications and $41$,$609$ citations between~them (\href{http://vlado.fmf.uni-lj.si}{{\scriptsize\tt http://vlado.fmf.uni-lj.si}}).
Note that $4$,$519$ publications neither cite nor are cited by any other, while the database includes $14$~self-citations.

\paragraph{\DBLP database.} 

\dblp (\DBLP) indexes major journals and proceedings from all fields of computer science~\cite{Ley02} (\href{http://dblp.uni-trier.de}{{\scriptsize\tt http://dblp.uni-trier.de}}). It is freely available since $1993$ and hand-maintained by University of Trier. It contains more than $2.3$ million records of publications, while the citation information is extremely scarce compared to \WoS and \CS databases~\cite{Fia11}. For this study, we consider a snapshot of the database provided within KONECT (\href{http://konect.uni-koblenz.de}{{\scriptsize\tt http://konect.uni-koblenz.de}}) that contains $12$,$591$ journal and conference papers, and $49$,$759$ citations between them.
Note that all papers either cite or are cited by some other, while
the database includes $15$~self-citations.

\paragraph{\arXiv database.} 

\arxiv (\arXiv) is a public preprint repository of publication drafts uploaded by the authors prior to an actual journal or conference submission (\href{http://arxiv.org}{{\scriptsize\tt http://arxiv.org}}). It began in $1991$~\cite{Gin11} and is hosted at Cornell University. It currently contains almost one million publications from physics, mathematics, computer science and other fields. For this study, we consider all publications in \arXiv category {\it High Energy Physics Phenomenology} as of April $2003$~\cite{LKF07} provided within SNAP (\href{http://snap.stanford.edu}{{\scriptsize\tt http://snap.stanford.edu}}). The database spans over $10$ years, and contains $34$,$546$ publications and $421$,$578$ citations between them.
Note that all publications either cite or are cited by some other, while
the database includes $44$~self-citations.

\subsection*{\label{sec:methods:statistics}Citation topology}

Citation networks extracted from bibliographic databases are represented with directed graphs, where papers are nodes of the graph and citations are directed links between the nodes. The topology of citation networks is assessed through a rich set of local and global graph statistics.

\paragraph{Descriptive and field statistics.} 

The citation network is a simple directed graph $\G(\V,\E)$, where $\V$ is the set of nodes, $\n=|\V|$, and $\E$ is the set of links, $\m=|\E|$. Weakly connected component (\WCC) is a subset of nodes reachable from one another not considering the directions of the links. Field bow-tie is a decomposition of the largest \WCC of a citation network into the in-field component, which consists of nodes with no outgoing links, the out-field component, which consists of nodes with no incoming links, and the field core.

\paragraph{Degree distributions and mixing.} 

The in-degree $\ki$ or out-degree $\ko$ of a node is the number of incoming and outgoing links, respectively. $\kd$ is the degree of a node, $\kd=\ki+\ko$, and $\avg{\kd}$ denotes the mean degree. $\plc$~is the scale-free exponent of a power-law degree distribution $P(\kd)\sim\kd^{-\plc}$, and $\plci$ and~$\plco$ are the scale-free exponents of $P(\ki)$ and $P(\ko)$~\cite{BA99}. Power-laws are fitted to the tails of the distributions by maximum-likelihood estimation, $\plc_{\cdot}=1+\n\left(\sum_{\V}\ln\left.\kd_{\cdot}/\kmin\right.\right)^{-1}$ for $\kmin\in\{10,25\}$. Neighbour connectivity plots show the mean neighbour degree $\N{\kd_{\cdot}}$ of nodes with degree $\kd_{\cdot}$~\cite{PVV01}. The~degree mixing $\mix_{(\alpha,\beta)}$ is the Pearson correlation coefficient of $\alpha$-degrees or $\beta$-degrees at links' source and~target nodes,~respectively~\cite{FFGP10}:
\begin{equation}\label{eq:r}
\mix_{(\alpha,\beta)} = \frac{1}{\std{\kd_{\alpha}}\std{\kd_{\beta}}}\sum_{\E}\left(\kd_{\alpha}-\avg{\kd_{\alpha}}\right)\left(\kd_{\beta}-\avg{\kd_{\beta}}\right),
\end{equation}
where $\avg{\kd_{\cdot}}$ and $\std{\kd_{\cdot}}$ are the means and standard deviations, $\alpha,\beta\in\{in,out\}$. $\mix$ is the mixing of degrees $\kd$~\cite{New03a}.

\paragraph{Clustering distributions and mixing.}

Node clustering coefficient $\clus$ is the density of its neighbourhood~\cite{WS98}:
\begin{equation}\label{eq:c}
\clus = \frac{2\trian}{\kd(\kd-1)},
\end{equation}
where $\trian$ is the number of linked neighbours and $\kd(\kd-1)/2$ is the maximum possible number, $\clus=0$ for $\kd\leq 1$. The mean $\avg{\clus}$ is denoted network clustering coefficient~\cite{WS98}, while the clustering mixing $\mixc$ is defined as before. Clustering profile shows the mean clustering $\C{\kd}$ of nodes with degree $\kd$~\cite{RSMOB02}. Note that the denominator in~\eqref{c} introduces biases~\cite{SV05}, particularly when $\mix<0$. Thus, delta-corrected clustering coefficient~$\bclus$ is defined as $\clus\cdot\kd/\h$~\cite{NMB05}, where $\h$ is the maximal degree $\kd$ and $\bclus=0$ for $\kd\leq 1$. Also, degree-corrected clustering coefficient $\dclus$ is defined as $\trian/\omg$~\cite{SV05}, where $\omg$ is the maximum number of linked neighbours with respect to their degrees $\kd$ and $\dclus=0$ for $\kd\leq 1$. By~definition, $\bclus\leq\clus\leq\dclus$.


\paragraph{Diameter statistics.} 

Hop plot shows the percentage of reachable pairs of nodes $\HP{\diam}$ within $\diam$ hops~\cite{LKF07}.
 The diameter is the minimal number of hops $\diam$ for which $\HP{\diam}=1$, while the effective diameter $\ediam$ is defined as the number of hops at which $90\%$ of such pairs of nodes are reachable~\cite{LKF07}, $\HP{\ediam}=0.9$. $\und{\diam}$ denotes the respective number of hops in a corresponding undirected graph. Hop plots are estimated over $100$ realizations of the approximate neighbourhood function with $32$ trials~\cite{PGF02}.

\subsection*{\label{sec:methods:comparison}Statistical comparison}

Citation networks representing bibliographic databases are compared through $21$ graph statistics introduced above. These are by no means independent~\cite{WS98,SV05}, neither are their values of a {\it true} citation network known. We thus compute externally studentized residuals of graph statistics that measure the consistency of each bibliographic database with the rest. Statistically significant inconsistencies in individual graph statistics are revealed by Student $t$-test~\cite{CW82}. We select ten graph statistics whose pairwise independence is verified using Fisher $z$-trans\-for\-mation~\cite{Fis15}. Friedman rank test~\cite{Fri37} confirms that bibliographic databases display significant inconsistencies in the selected statistics, while the databases with no significant differences are revealed by Nemenyi test~\cite{Nem63,Dem06}. 

\paragraph{Studentized statistics residuals.} 

Denote $\x_{ij}$ to be the value of $j$-th graph statistic of $i$-th bibliographic~data\-base, where $\nets$ is the number of databases, $\nets=6$. Corresponding externally studentized residual $\esr{\x}_{ij}$~is:
\begin{equation}\label{eq:x}
\esr{\x}_{ij}=\frac{\x_{ij}-\esr{\mean{}}_{ij}}{\esr{\std{}}_{ij}\sqrt{1-1/\nets}},
\end{equation}
where $\esr{\mean{}}_{ij}$ and $\esr{\std{}}_{ij}$ are the sample mean and corrected standard deviation excluding the considered $i$-th database, $\esr{\mean{}}_{ij}=\sum_{k\neq i}\x_{kj}/(\nets-1)$ and $\esr{\std{}}_{ij}^2=\sum_{k\neq i}(\x_{kj}-\esr{\mean{}}_{ij})^2/(\nets-2)$. Assuming that the errors in $\x$ are~independent and normally distributed, the residuals $\esr{\x}$ have Student $t$-distribution with $\nets-2$ degrees of freedom. Significant differences in individual statistics $\x$ are revealed by independent two-tailed Student $t$-tests~\cite{CW82} at $\pval=0.05$, rejecting the null hypothesis $\nhyp$ that $\x$ are consistent across the databases, $\nhyp:\esr{\x}=0$. Notice that the absolute values of individual residuals $|\esr{\x}|$ imply a ranking $\rank$ over the databases, where the database with the lowest $|\esr{\x}|$ has rank one, the second one has rank two and the one with the largest $|\esr{\x}|$ has rank $\nets$.

\paragraph{Pairwise statistics independence.}

Denote $\pcc_{ij}$ to be the Pearson product-moment correlation coefficient of the residuals $\esr{\x}$ for $i$-th and $j$-th graph statistics over all bibliographic databases. Spearman rank correlation coefficient $\scc_{ij}$ is defined as the Pearson coefficient of the ranks $\rank$ for $i$-th and $j$-th statistics. Under the null hypothesis of statistical independence of $i$-th and $j$-th statistics, $\nhyp:\scc_{ij}=0$, adjusted Fisher~transformation~\cite{Fis15}:
\begin{equation}\label{eq:z}
\frac{\sqrt{\nets-3}}{2}\ln\left.\frac{1+\pcc_{ij}}{1-\pcc_{ij}}\right.
\end{equation}
approximately follows a standard normal distribution. Pairwise independence of the selected graph statistics is thus confirmed by independent two-tailed $z$-tests at $\pval=0.01$.

\paragraph{Comparison of bibliographic databases.}

Significant inconsistencies between bibliographic databases are exposed using the methodology introduced for comparing classification algorithms over multiple data sets~\cite{Dem06}. Denote $\rank_i$ to be the mean rank of $i$-th database over the selected graph statistics, $\rank_i=\sum_j\rank_{ij}/\stats$, where $\stats$ is the number of statistics, $\stats=10$. One-tailed Friedman rank test~\cite{Fri37,Fri40} first verifies the null hypothesis that the databases are statistically equivalent and thus their ranks $\rank_i$ should equal, $\nhyp:\rank_i=\rank_j$. Under the assumption that the selected statistics are indeed independent, the Friedman testing statistic~\cite{Fri37}:
\begin{equation}\label{eq:F}
\frac{12\stats}{\nets(\nets+1)}\left(\sum_i\rank_i^2-\frac{\nets(\nets+1)^2}{4}\right)
\end{equation}
has $\chi^2$-distribution with $\nets-1$ degrees of freedom. By rejecting the hypothesis at $\pval=0.05$, we proceed with the Nemenyi post-hoc test that reveals databases whose ranks $\rank_i$ differ more than the critical~difference~\cite{Nem63}:
\begin{equation}\label{eq:cd}
\cv\sqrt{\frac{\nets(\nets+1)}{6\stats}},
\end{equation}
where $\cv$ is the critical value based on the studentized range statistic~\cite{Dem06}, $\cv=2.85$ at $\pval=0.05$. A critical difference diagram plots the databases with no statistically significant inconsistencies in the selected statistics~\cite{Dem06}.

}


{\footnotesize

\bibliographystyle{naturemag}

}


\section*{Acknowledgements}
Authors thank J. Dem\v{s}ar, V.\ Batagelj, M.\ \v{Z}itnik and Z.\ Levnaji\'{c} for comments and discussions, and Thomson Reuters for providing the access to bibliographic data. This work has been supported in part by the Slovenian Research Agency Program No.\ P2-0359, by the Slovenian Ministry of Education, Science and Sport Grant No.\ 430-168/2013/91, by the European Union, European Social Fund, and by the European Regional Development Fund Grant No.\ CZ.1.05/1.1.00/02.0090.

\section*{Author Contributions}
L.\v{S}. designed and performed the experiments. L.\v{S}., D.F. and M.B. wrote the main manuscript text. All authors reviewed the manuscript. The authors have no competing financial interests.

\section*{Additional Information}
The authors declare no competing financial interests.

\clearpage

\pagestyle{empty}


\begin{center}
  {\huge Figure Legends}
\end{center}

\begin{figure}[!h]
	\captionsetup{font=small}
	\caption{\label{fig:profile}{\bf Profile of citation networks extracted from bibliographic~databases.} Panels {\bf A-F} show different distributions, plots and profiles of citation networks extracted from bibliographic databases. These are (from left to right): the field bow-tie decompositions, where the arrows illustrate the direction of the links and the areas of components are proportional to the number of nodes contained; the degree, in-degree and out-degree distributions $P(\kd)$, $P(\ki)$ and $P(\ko)$, respectively; the corresponding neighbour connectivity plots $\N{\kd}$, $\N{\ki}$ and $\N{\ko}$; the clustering profiles of the standard and both unbiased coefficients $\C{\kd}$, $\B{\kd}$ and $\D{\kd}$, respectively; and the hop plots for the standard and undirected diameters $\diam$ and $\und{\diam}$, respectively (see~\metref).}
\end{figure}

\begin{figure}[!h]
	\captionsetup{font=small}
	\caption{\label{fig:comparison}{\bf Comparison of bibliographic databases through statistics of citation networks.} Panels {\bf A-F} show studentized statistics residuals of citation networks extracted from bibliographic databases. The residuals are listed in decreasing order, while the shaded regions
are $95\%$ and $99\%$ confidence intervals of independent Student $t$-tests (labelled with respective $\pval$s). Panel {\bf G} shows the residuals of merely independent statistics, where the shaded region is $95\%$ confidence interval. Panel {\bf H} shows pairwise Spearman correlations of independent statistics listed in the same order as in panel {\bf G} (left) and the $\pval$s of the corresponding Fisher independence $z$-tests (right). Panel {\bf I} shows the critical difference diagram of Nemenyi post-hoc test for the independent statistics. The diagram illustrates the overall ranking of the databases, where those connected by a thick line show no statistically significant inconsistencies at $\pval=0.05$ (see~\metref).}
\end{figure}

\begin{figure}[!h]
	\captionsetup{font=small}
	\caption{\label{fig:validation}{\bf Comparison of bibliographic and online databases through statistics of networks.} Panels {\bf A-D} show studentized statistics residuals of citation networks extracted from bibliographic databases, while panels {\bf E} and {\bf F} show residuals of social and technological networks extracted from online databases. The residuals are listed in decreasing order, while the shaded regions are $95\%$ and $99\%$ confidence intervals of independent Student $t$-tests (labelled with respective $\pval$s). Panel {\bf G} shows the residuals of merely independent statistics, where the shaded region is $95\%$ confidence interval. Panel {\bf H} shows pairwise Spearman correlations of independent statistics listed in the same order as in panel {\bf G} (left) and the $\pval$s of the corresponding Fisher independence $z$-tests (right). Panel {\bf I} shows the critical difference diagram of Nemenyi post-hoc test for the independent statistics. The diagram illustrates the overall ranking of the databases, where those connected by a thick line show no statistically significant inconsistencies at $\pval=0.05$ (see~\metref).}
\end{figure}


\clearpage


\begin{center}
  {\huge Tables}
\end{center}

\rowcolors{1}{gray!1}{gray!20}
\begin{table}[!h] \centering\small
	\begin{tabular}{lcccccc} \hline
		& \multicolumn{3}{c}{Descriptive statistics} & \multicolumn{3}{c}{Field decomposition} \\\hline
		Source & \# Nodes & \# Links & \% WCC & \% In-field & \% Core & \% Out-field \\\hline
		\WoS & $140$,$362$ & $\phantom{0,}639$,$110$ & $\phantom{0}97.0\%$ & $\phantom{0}11.2\%$ & $\phantom{0}51.4\%$ & $\phantom{0}34.4\%$ \\
		\CS & $384$,$413$ & $1$,$744$,$619$ & $\phantom{0}95.0\%$ & $\phantom{0}10.5\%$ & $\phantom{0}37.7\%$ & $\phantom{0}46.8\%$ \\
		\Cora & $\phantom{0}23$,$166$ & $\phantom{0,0}91$,$500$ & $100.0\%$ & $\phantom{00}8.5\%$ & $\phantom{0}51.4\%$ & $\phantom{0}40.1\%$ \\
		\HC & $\phantom{00}4$,$324$ & $\phantom{0,0}41$,$595$ & $\phantom{0}98.7\%$ & $\phantom{0}44.8\%$ & $\phantom{0}52.2\%$ & $\phantom{00}1.6\%$ \\
		\DBLP & $\phantom{0}12$,$591$ & $\phantom{0,0}49$,$744$ & $\phantom{0}99.2\%$ & $\phantom{0}74.5\%$ & $\phantom{0}16.9\%$ & $\phantom{00}7.8\%$ \\
		\arXiv & $\phantom{0}34$,$546$ & $\phantom{0,}421$,$534$ & $\phantom{0}99.6\%$ & $\phantom{00}6.7\%$ & $\phantom{0}74.7\%$ & $\phantom{0}18.1\%$ \\\hline
		\Gnutella & $\phantom{0}62$,$586$ & $\phantom{0,}147$,$892$ & $100.0\%$ & $\phantom{0}73.8\%$ & $\phantom{0}25.7\%$ & $\phantom{00}0.5\%$ \\
		\Twitter & $\phantom{0}81$,$306$ & $1$,$768$,$135$ & $100.0\%$ & $\phantom{0}13.8\%$ & $\phantom{0}86.2\%$ & $\phantom{00}0.0\%$ \\\hline
	\end{tabular}
	\captionsetup{font=small}
	\caption{\label{tbl:statistics}{\bf Descriptive statistics and field decompositions of citation and other networks.} Respective bibliographic or online data\-bases are given under the column denoted by \col{Source}. Descriptive statistics list the number of network nodes~$\n$ and links $\m$, and the percentage of nodes in the largest weakly connected component (column labelled \col{\% WCC}). Columns labelled \col{\% In-field}, \col{\% Core} and \col{\% Out-field} report the percentages of nodes in each of the components of the field bow-tie decomposition (see~\metref).}
\end{table}

\rowcolors{1}{gray!1}{gray!20}
\begin{table}[!h] \centering\small
	\begin{tabular}{lccccccccc} \hline
		& \multicolumn{4}{c}{Degree distributions} & \multicolumn{5}{c}{Degree mixing} \\\hline
		Source & $\avg{\kd}$ & $\plc$ & $\plci$ & $\plco$ & $\mix$ & $\mixii$ & $\mixio$ & $\mixoi$ & $\mixoo$ \\\hline
		\WoS & $\phantom{0}9.11$ & $\phantom{0}2.74$ & $\phantom{0}2.39$ & $\phantom{0}3.88$ & $-0.06$ & $\phantom{-}0.04$ & $-0.02$ & $-0.03$ & $\phantom{-}0.09$ \\
		\CS & $\phantom{0}9.08$ & $\phantom{0}2.65$ & $\phantom{0}2.28$ & $\phantom{0}3.82$ & $-0.06$ & $\phantom{-}0.05$ & $\phantom{-}0.00$ & $\phantom{-}0.00$ & $\phantom{-}0.12$ \\
		\Cora & $\phantom{0}7.90$ & $\phantom{0}2.88$ & $\phantom{0}2.60$ & $\phantom{0}4.00$ & $-0.06$ & $\phantom{-}0.07$ & $\phantom{-}0.02$ & $\phantom{-}0.00$ & $\phantom{-}0.17$ \\
 		\HC & $\phantom{0}9.99$ & $\phantom{0}2.55$ & $\phantom{0}3.50$ & $\phantom{0}2.37$ & $-0.10$ & $\phantom{-}0.11$ & $\phantom{-}0.01$ & $-0.13$ & $\phantom{-}0.00$ \\
		\DBLP & $\phantom{0}7.90$ & $\phantom{0}2.42$ & $\phantom{0}2.64$ & $\phantom{0}2.75$ & $-0.05$ & $\phantom{-}0.00$ & $-0.02$ & $-0.05$ & $-0.02$ \\
		\arXiv & $24.40$ & $\phantom{0}2.67$ & $\phantom{0}2.54$ & $\phantom{0}3.45$ & $-0.01$ & $\phantom{-}0.08$ & $-0.04$ & $\phantom{-}0.00$ & $\phantom{-}0.11$ \\\hline
		\Gnutella & $\phantom{0}4.73$ & $\phantom{0}6.37$ & $\phantom{0}7.59$ & $\phantom{0}4.78$ & $-0.09$ & $\phantom{-}0.03$ & $\phantom{-}0.01$ & $-0.01$ & $\phantom{-}0.00$ \\
		\Twitter & $43.49$ & $\phantom{0}2.05$ & $\phantom{0}2.31$ & $\phantom{0}2.37$ & $-0.03$ & $\phantom{-}0.00$ & $\phantom{-}0.06$ & $-0.02$ & $\phantom{-}0.06$ \\\hline
	\end{tabular}
	\captionsetup{font=small}
	\caption{\label{tbl:degree}{\bf Degree distributions and mixing of citation and other networks.} Respective bibliographic or online data\-bases are given under the column denoted by \col{Source}. Degree distributions are represented by the mean network degree $\avg{\kd}$ and the scale-free exponents of the power-law degree, in-degree and out-degree distributions (columns labelled \col{$\plc$}, \col{$\plci$} and \col{$\plco$}, respectively). Degree mixing statistics list the undirected mixing coefficient $\mix$ and four directed degree mixing coefficients $\mix_{(\alpha,\beta)}$, $\alpha,\beta\in\{in,out\}$ (see~\metref).}
\end{table}

\rowcolors{1}{gray!1}{gray!20}
\begin{table}[!h] \centering\small
	\begin{tabular}{lcccccccc} \hline
		& \multicolumn{3}{c}{Clustering distributions} & \multicolumn{3}{c}{Clustering mixing} & \multicolumn{2}{c}{Diameter statistics} \\\hline
		Source & $\avg{\clus}$ & $\avg{\bclus}$ & $\avg{\dclus}$ & $\mixc$ & $\mixb$ & $\mixd$ & $\ediam$ & $\und{\ediam}$ \\\hline
		\WoS  & $\phantom{0}0.14$ & $\phantom{0}0.08\cdot 10^{-2}$ & $\phantom{0}0.16$ & $\phantom{0}0.16$ & $\phantom{0}0.43$ & $\phantom{0}0.36$ &  $\phantom{0}8.85\pm0.01$ & $\phantom{0}7.79\pm0.03$ \\
		\CS & $\phantom{0}0.18$ & $\phantom{0}0.07\cdot 10^{-2}$ & $\phantom{0}0.21$ & $\phantom{0}0.14$ & $\phantom{0}0.44$ & $\phantom{0}0.40$ & $28.57\pm0.23$ & $\phantom{0}9.01\pm0.04$ \\
		\Cora & $\phantom{0}0.27$ & $\phantom{0}0.46\cdot 10^{-2}$ & $\phantom{0}0.32$ & $\phantom{0}0.17$ & $\phantom{0}0.50$ & $\phantom{0}0.40$ & $21.12\pm0.16$ & $\phantom{0}8.17\pm0.03$ \\
		\HC & $\phantom{0}0.31$ & $\phantom{0}0.20\cdot 10^{-2}$ & $\phantom{0}0.36$ & $\phantom{0}0.05$ & $\phantom{0}0.36$ & $\phantom{0}0.41$ & $\phantom{0}7.97\pm0.03$ & $\phantom{0}7.22\pm0.04$ \\
		\DBLP & $\phantom{0}0.12$ & $\phantom{0}0.14\cdot 10^{-2}$ & $\phantom{0}0.14$ & $\phantom{0}0.10$ & $\phantom{0}0.35$ & $\phantom{0}0.26$ & $\phantom{0}9.13\pm0.07$ & $\phantom{0}6.24\pm0.02$ \\
		\arXiv & $\phantom{0}0.28$ & $\phantom{0}0.64\cdot 10^{-2}$ & $\phantom{0}0.33$ & $\phantom{0}0.13$ & $\phantom{0}0.46$ & $\phantom{0}0.39$ & $21.71\pm0.12$ & $\phantom{0}6.04\pm0.02$ \\\hline
		\Gnutella& $\phantom{0}0.01$ & $\phantom{0}0.03\cdot 10^{-2}$ & $\phantom{0}0.01$ & $\phantom{0}0.09$ & $\phantom{0}0.25$ & $\phantom{0}0.17$ & $12.83\pm0.11$ & $\phantom{0}7.70\pm0.01$ \\
		\Twitter & $\phantom{0}0.57$ & $\phantom{0}0.35\cdot 10^{-2}$ & $\phantom{0}0.63$ & $\phantom{0}0.09$ & $\phantom{0}0.54$ & $\phantom{0}0.40$ & $\phantom{0}6.90\pm0.02$ & $\phantom{0}5.50\pm0.01$ \\\hline
	\end{tabular}
	\captionsetup{font=small}
	\caption{\label{tbl:clustering}{\bf Clustering and diameter statistics of citation and other networks.} Respective bibliographic or online databases are given under the column denoted by \col{Source}. Clustering distributions are represented by the~means of the standard and unbiased clustering coefficients (columns labelled \col{$\avg{\clus}$}, \col{$\avg{\bclus}$} and \col{$\avg{\dclus}$}, respectively). Clustering mixing statistics list the corresponding mixing coefficients $\mixc$, $\mixb$ and $\mixd$. Diameter statistics report the means and s.e.m.\ of the standard and undirected effective diameters (columns labelled \col{$\ediam$} and \col{$\und{\ediam}$},~respectively).}
\end{table}

\clearpage


\begin{center}
  {\huge Figures}
\end{center}

\rowcolors{1}{}{}
\begin{figure}[!h] \centering\small
\includegraphics[width=\textwidth]{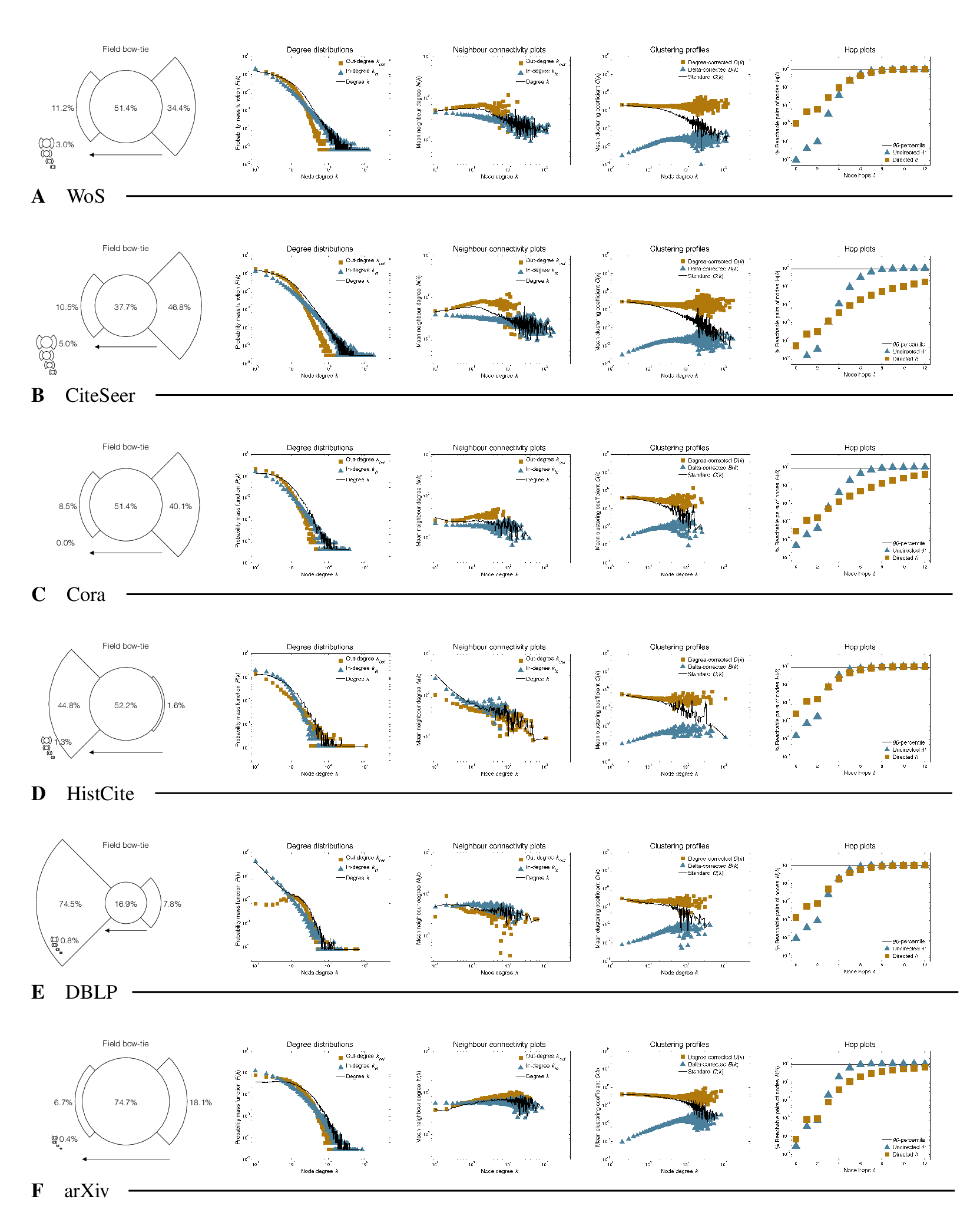}
	\captionsetup{font=small}
	\caption*{\sfigref{profile}.}
\end{figure}

\rowcolors{1}{}{}
\begin{figure}[!h] \centering\small
\includegraphics[width=\textwidth]{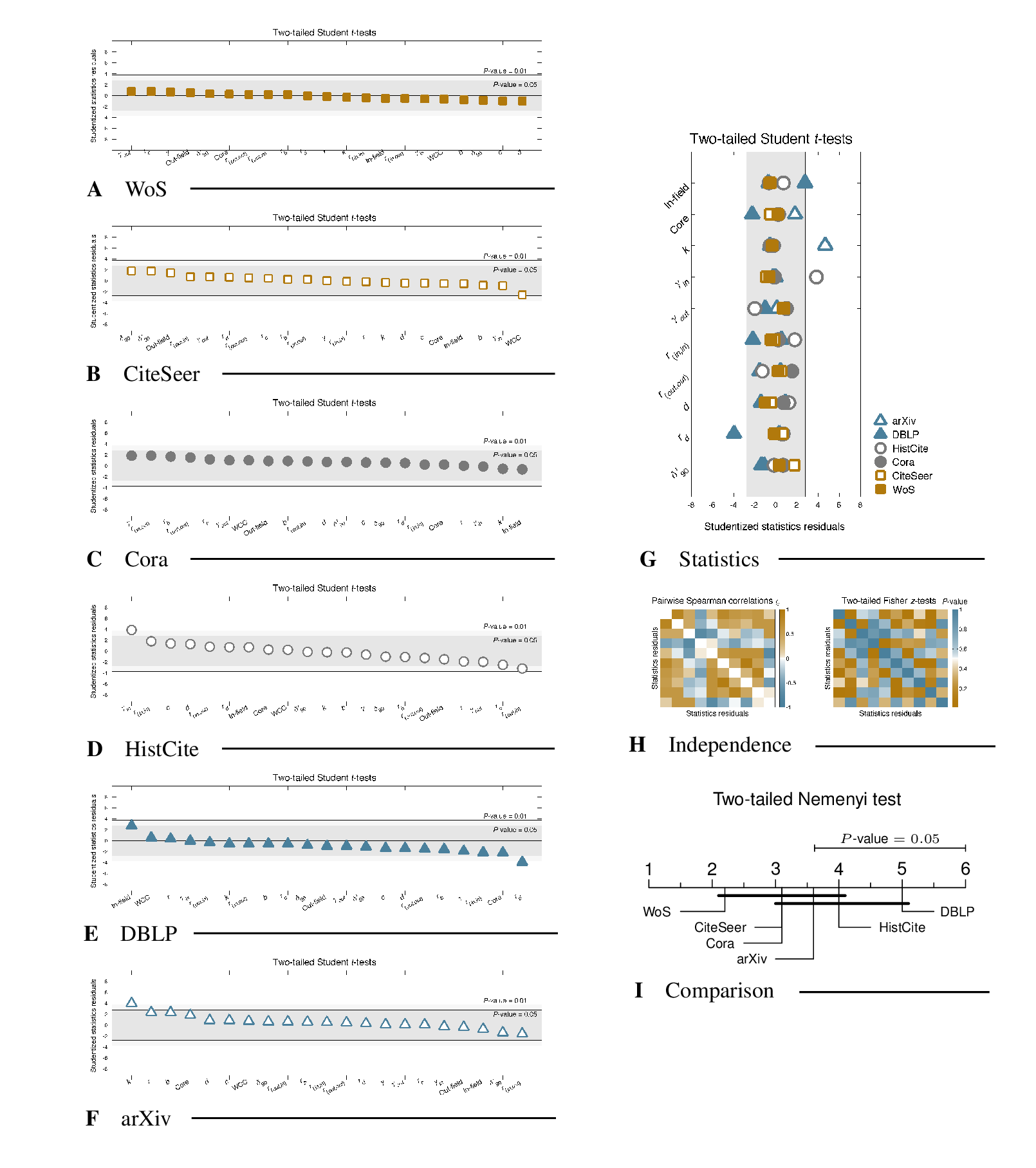}
	\captionsetup{font=small}
	\caption*{\sfigref{comparison}.}
\end{figure}

\rowcolors{1}{}{}
\begin{figure}[!h] \centering\small
\includegraphics[width=\textwidth]{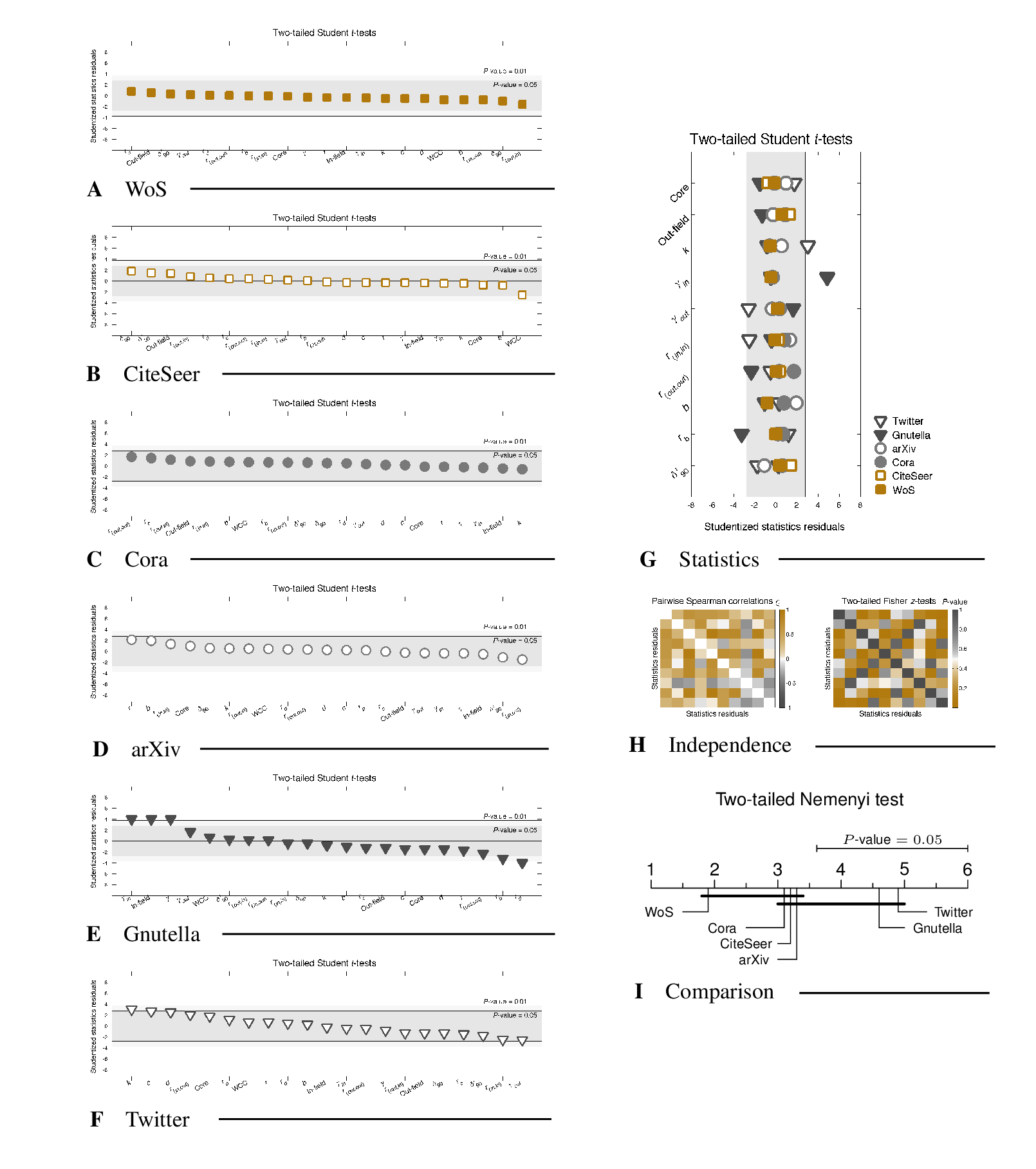}
	\captionsetup{font=small}
	\caption*{\sfigref{validation}.}
\end{figure}

\clearpage


\end{document}